\documentclass[12pt]{article}
\usepackage{amsmath,amsfonts,feynmp,epsf}
\usepackage{amssymb}
\usepackage{graphicx}
\usepackage{grffile}
\input epsf

\makeatletter\@addtoreset{equation}{section}\makeatother

\def\be{\begin{equation}}
\def\ee{\end{equation}}
\def\bea{\begin{eqnarray}}
\def\eea{\end{eqnarray}}
\def\ie{\begin{equation}\begin{aligned}}
\def\fe{\end{aligned}\end{equation}}

\newcommand{\A}{{\alpha}}

\makeatletter\@addtoreset{equation}{section}\makeatother

\renewcommand{\title}[1]{\vbox{\center\LARGE{#1}}\vspace{5mm}}
\renewcommand{\author}[1]{\vbox{\center#1}\vspace{5mm}}
\newcommand{\address}[1]{\vbox{\center\em#1}}
\newcommand{\email}[1]{\vbox{\center\tt#1}\vspace{5mm}}

\begin{document}
\unitlength = .8mm

\begin{titlepage}

\begin{center}
\hfill \\
\hfill \\
\vskip 1cm
\title{On Higher Spin Gauge Theory \\ and the Critical $O(N)$ Model}


\author{Simone Giombi$^{1,a}$ and
Xi Yin$^{2,b}$}

\address{${}^1$Perimeter Institute for Theoretical Physics, Waterloo, Ontario, N2L 2Y5, Canada}
\address{
${}^2$Center for the Fundamental Laws of Nature,
Jefferson Physical Laboratory,\\
Harvard University,
Cambridge, MA 02138 USA}

\email{$^a$sgiombi@pitp.ca, 
$^b$xiyin@fas.harvard.edu}

\end{center}

\abstract{We show that the differences between correlators of the critical $O(N)$ vector model in three dimensions and those of the free theory are precisely accounted for by the change of boundary condition on the bulk scalar of the dual higher spin gauge theory in $AdS_4$. Thus, the conjectured duality between Vasiliev's theory and the critical $O(N)$ model follows, order by order in $1/N$, from the duality with free field theory on the boundary.}

\vfill

\end{titlepage}

\section{Introduction}

One of the simplest nontrivial examples of the AdS/CFT correspondence \cite{Maldacena:1997re} is the conjectured duality \cite{Klebanov:2002ja, Sezgin:2002rt} between Vasiliev's higher spin gauge theory in $AdS_4$ \cite{Vasiliev:1999ba} and $O(N)$ vector models. At the classical level, Vasiliev's system gives a set of nonlinear equations of motion for an infinite set of gauge fields of spins $s=2,4,6,\ldots$ and a scalar field with $m^2=-2/R_{AdS}^2$.\footnote{This is the spectrum of the so-called ``minimal" bosonic Vasiliev's theory. It is a consistent truncation of the more general ``non-minimal" system, which also includes all odd spins. The dual of the non-minimal theory is expected to be a $U(N)$ vector model, restricted to the $U(N)$ singlet sector.}  The mass is precisely in the window which allows a choice of two different boundary conditions on the bulk scalar field $\varphi$, such that the dual operator has classical dimension $\Delta=1$ or $\Delta=2$. The bulk theory with $\Delta=1$ boundary condition is conjectured to be dual to the free theory of $N$ massless scalars $\phi_i$ in three dimensions, restricted to the $O(N)$ singlet sector, whereas the bulk theory with $\Delta=2$ boundary condition is conjectured to be dual to the critical $O(N)$ vector model, which may be described by the critical point of the $S^{N-1}$ non-linear $\sigma$-model with Lagrangian 
\ie\label{lagr}
{\cal L} ={N\over 2}\left[ (\partial_\mu \phi_i)^2 + \tilde\A \big(\phi_i \phi_i - {1\over g}\big)\right].
\fe
Here $\tilde\A$ is a Lagrange multiplier field, and the critical point is achieved by sending $g\to \infty$. A systematic $1/N$ expansion of the critical $O(N)$ model has been studied in \cite{LangRuhl, Petkou:1994ad}. Alternatively, the critical theory may be defined as the IR fixed point of a relevant $(\phi_i \phi_i)^2$ deformation of the free theory.

In principle, the bulk Vasiliev's theory is computable perturbatively, which corresponds to the $1/N$ expansion of the dual $O(N)$ vector model. The first such perturbative computation was carried out in \cite{Giombi:2009wh, Giombi:2010vg}, and highly nontrivial agreement of three point functions between the bulk and boundary theories have been found at leading order in $1/N$, for both $\Delta=1$ and $\Delta=2$ boundary conditions (see \cite{Sezgin:2002ru, Sezgin:2003pt, Petkou:2003zz, Leigh:2003gk} for earlier works, and \cite{Koch:2010cy, Douglas:2010rc} for some new perspectives). 

Eventually, one would like to compute all the $n$-point functions from the bulk theory, and have a perturbative proof of the duality. While the agreement between Vasiliev's system with $\Delta=1$ boundary condition and the {\it free} $O(N)$ theory may not be surprising, given that the free theory is our only known example of CFTs in dimension greater than two with exactly conserved higher spin currents, the duality in the case of $\Delta=2$ boundary condition, which breaks higher spin symmetry in the bulk through loop effects, has been more mysterious (see \cite{Girardello:2002pp, Manvelyan:2008ks} for earlier work on this mechanism). This is perhaps also the more interesting case as the dual CFT is an interacting theory. 

In this paper, we will address the duality in the case of $\Delta=2$ boundary condition. Thanks to a simple factorization identity involving the bulk scalar propagators for the two different boundary conditions, we will give a perturbative argument that the {\it difference} between correlators in the $\Delta=2$ and $\Delta=1$ theories as computed from the bulk theory precisely accounts for the difference between those corresponding correlators in the critical $O(N)$ vector model and the free theory. The duality in the $\Delta=2$ case, to all order in $1/N$, then follows from the duality in the $\Delta=1$ case where the higher spin symmetry is preserved. This also clarifies and confirms the breaking of higher spin symmetry through loops of bulk scalars, which gives a finite mass renormalization of the bulk higher spin fields through its mixing with two-particle states involving a higher spin field and a scalar \cite{Girardello:2002pp}. In some sense our arguments are an extension of the Legendre transform relating the two boundary conditions \cite{Witten:2001ua, Gubser:2002vv} to all order in $1/N$.

We now begin with the simple examples of tree level three and four point functions, which illustrates our argument, and then discuss the general $n$ point functions and loop corrections.

Note added:\footnote{We thank L. Rastelli for pointing this out to us.} Upon completion of this paper, we were informed that the results of section 4 and the key observation of the factorization of the difference between $\Delta=1$ and $\Delta=2$ bulk scalar propagators in momentum space have already appeared in \cite{Hartman:2006dy}.

\section{Three-point functions with a scalar operator}

The ``single-trace" primary operators in the critical $O(N)$ vector model are the currents $J^{(s)}$, $s=2,4,\ldots$ of dimension $\Delta=s+1+{\cal O}\left(\frac{1}{N}\right)$, and the scalar Lagrange multiplier field $\alpha$ with $\Delta=2+{\cal O}\left(\frac{1}{N}\right)$.\footnote{Here and later on, $\A$ will be normalized by a canonical normalization on its two-point function, which differs from that of $\tilde\A$ in (\ref{lagr}).} Let $J^{(s)}_{\mu_1\cdots\mu_s}$ be the spin-$s$ current. By definition it is symmetric and traceless in $(\mu_1,\cdots,\mu_s)$, though not conserved for $s>2$ at finite $N$. It can be expressed in terms of the fundamental scalar fields $\phi_i$ as 
\ie\label{curr}
J^{(s)}(x,\varepsilon) \equiv J^{(s)}_{\mu_1\cdots\mu_s}(x) \varepsilon^{\mu_1}\cdots\varepsilon^{\mu_s}
= \phi_i f(\varepsilon\cdot \overleftarrow \partial, \varepsilon\cdot \overrightarrow \partial) \phi_i 
\fe
where $\varepsilon^\mu$ is an arbitrary null polarization vector, and the function $f(u,v)$ is given by
\ie
f(u,v) = e^{u-v} \cos(2\sqrt{u v}).
\label{fuv}
\fe
The precise form of $f(u,v)$ will not be needed in what follows. Note that (\ref{curr}),(\ref{fuv}) is the free field expression for the higher spin currents \cite{Giombi:2009wh}, which also holds in the critical $O(N)$ theory. This is because $J^{(s)}$ has classical dimension $\Delta=s+1$ and cannot mix with multi-trace operators (which have $\Delta-s\geq 2$) or operators that involve $\alpha$ (the scalar operator of classical dimension 2), and so (\ref{curr}) is the correct expression for the spin $s$ primary operator in the critical theory. In particular, it guarantees that $\langle J^{(s)}\, \alpha\rangle=0$.

Now consider the three point function $\langle \A(x_1) J(x_2) J'(x_3)\rangle$, where $J$ and $J'$ are two higher spin operators. At leading order in $1/N$, in momentum space, this is given by the corresponding three point function $\langle {\cal O}(p) J(q) J'(-p-q)\rangle$ in the {\it free} $O(N)$ theory (here ${\cal O}=\phi^i\phi^i$ is the $\Delta=1$ scalar operator), multiplied by the propagator for $\A$, 
\ie
D_\A(p) = \langle\A(p)\A(-p)\rangle = - |p|.
\fe
In the bulk, $\A$ is dual to the scalar field $\varphi$ with boundary to bulk propagator 
\ie
K_{\Delta}(x;\vec x_0) = {\Gamma(\Delta)\over \pi^{3\over 2} \Gamma(\Delta-{3\over 2})} \left[ {z\over (\vec x-\vec x_0)^2+z^2} \right]^\Delta
\fe
with $\Delta=2$. Its Fourier transform in $\vec x$ is
\ie
K_{\Delta=2}(p,z) =\int d^3x \,e^{i\vec{p}\cdot \vec{x}}K_{\Delta=2}(x;\vec x_0)=  z e^{-|p|z}.
\fe
Similarly, the momentum space boundary to bulk propagator for the scalar in the $\Delta=1$ case is
\ie
K_{\Delta=1}(p,z) = -{z\over |p|} e^{-|p|z},
\fe
and so
\ie
K_{\Delta=2}(p,z) = - |p| K_{\Delta=1}(p,z) .
\fe
Therefore, if we are to replace an external $\Delta=1$ scalar line of the Witten diagram by a
$\Delta=2$ scalar line, the resulting boundary correlator in momentum space is multiplied by a factor of $-|p|$ where $p$ is the momentum of the corresponding boundary scalar operator. This is precisely the correct relation between the correlators in the critical and free $O(N)$ vector models.

The bulk Witten diagrams and the boundary Feyman diagrams for this three point function are illustrated in the figures below (see \cite{LangRuhl, Giombi:2009wh} for detailed discussions of the $1/N$ expansion of the critical $O(N)$ model.)

\bigskip 

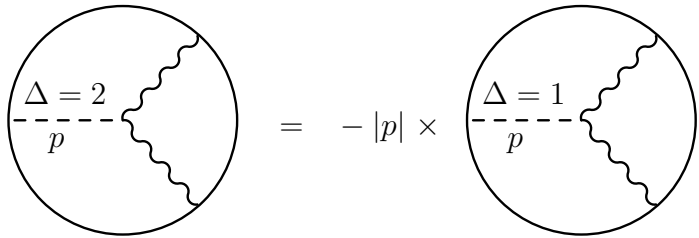
\centerline{\begin{fmffile}{OJJ}
        \begin{tabular}{c}
            \begin{fmfgraph*}(40,40)
            \fmfi{plain}{fullcircle scaled .95w shifted (.5w,.5h)}
                \fmfleft{L}
                \fmfright{R}
                \fmffixed{(-.05w,0)}{vO,L}
                \fmffixed{(.19w,0.35h)}{vJ1,R}
                \fmffixed{(.19w,-0.35h)}{vJ2,R}
                \fmffixed{(-.5w,0)}{v,L}
                \fmf{wiggly}{vJ1,v,vJ2}
                \fmf{dashes}{vO,v}
                \fmffixed{(.02w,-.02h)}{t,v}
                \fmffixed{(.16w,.09h)}{t2,t}
                \fmflabel{$\Delta=2$}{t}
                \fmflabel{$p$}{t2}
                \end{fmfgraph*}
        \end{tabular}
        \end{fmffile}
$=~~- |p|~\times\!\!\!$\begin{fmffile}{OJJ2}
        \begin{tabular}{c}
            \begin{fmfgraph*}(40,40)
            \fmfi{plain}{fullcircle scaled .95w shifted (.5w,.5h)}
                \fmfleft{L}
                \fmfright{R}
                \fmffixed{(-.05w,0)}{vO,L}
                \fmffixed{(.19w,0.35h)}{vJ1,R}
                \fmffixed{(.19w,-0.35h)}{vJ2,R}
                \fmffixed{(-.5w,0)}{v,L}
                \fmf{wiggly}{vJ1,v,vJ2}
                \fmf{dashes}{vO,v}
                \fmffixed{(.02w,-.02h)}{t,v}
                \fmflabel{$\Delta=1$}{t}
                \fmffixed{(.16w,.09h)}{t2,t}
                \fmflabel{$p$}{t2}
             \end{fmfgraph*}
        \end{tabular}
        \end{fmffile}
}
\bigskip
\centerline{Bulk tree level three-point function with $\Delta=2$ and $\Delta=1$ boundary conditions.}
\noindent

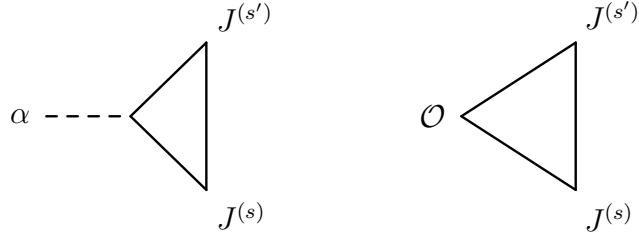
\centerline{\begin{fmffile}{OJJcrit}
        \begin{tabular}{c}
            \begin{fmfgraph*}(35,35)
                \fmfleft{L}
                \fmfright{R}
                \fmffixed{(-.05w,0)}{vO,L}
                \fmffixed{(.19w,0.35h)}{vJ1,R}
                \fmffixed{(.19w,-0.35h)}{vJ2,R}
                \fmffixed{(-.45w,0)}{v,L}
                \fmf{plain}{vJ1,v,vJ2,vJ1}
                \fmf{dashes}{vO,v}
                \fmflabel{$\alpha$}{vO}
                \fmflabel{$J^{(s)}$}{vJ1}
                \fmflabel{$J^{(s')}$}{vJ2}
             \end{fmfgraph*}
        \end{tabular}
        \end{fmffile}
~~~~~~~~~~~~        
        \begin{fmffile}{OJJfree}
        \begin{tabular}{c}
            \begin{fmfgraph*}(25,35)
                \fmfleft{L}
                \fmfright{R}
                \fmffixed{(-.05w,0)}{vO,L}
                \fmffixed{(.19w,0.35h)}{vJ1,R}
                \fmffixed{(.19w,-0.35h)}{vJ2,R}
                \fmffixed{(-.45w,0)}{v,L}
                \fmf{plain}{vJ1,vO,vJ2,vJ1}
                \fmflabel{${\cal O}$}{vO}
                \fmflabel{$J^{(s)}$}{vJ1}
                \fmflabel{$J^{(s')}$}{vJ2}
             \end{fmfgraph*}
        \end{tabular}
        \end{fmffile}
}
\bigskip
\centerline{The corresponding computation in the critical and free $O(N)$ vector models.}
\noindent

This agreement also indicates how higher spin symmetry is broken by the $\Delta=2$ boundary condition. While $\langle {\cal O} J J' \rangle$ clearly obeys the Ward identity due to the conservation of currents $J$ and $J'$, $\langle \A J J'\rangle$ generally violates such Ward identity even at leading order in $1/N$, when $J$ and $J'$ have different spins. This is because of the mixing of the divergence of the current with a double trace operator, (here $J^{(s)}$ is normalized such that its two point function does not scale with $N$)
\ie\label{mix}
\partial^\mu J^{(s)}_{\mu \mu_1\cdots\mu_{s-1}}(x) \sim {1\over \sqrt{N}} \sum_{n+m+s'=s-1} \partial^n \A(x)\, \partial^m J^{(s')}(x)
\fe
which may be derived by applying the classical equation of motion from the critical $O(N)$ model Lagrangian.\footnote{Such mixing between the divergence of the current and double trace operators was first pointed out to us by Shiraz Minwalla, and is explored in detail in \cite{GMPTY}.}
An explicit example is discussed in more detail in the appendix.

In the bulk computation, naively, the boundary-to-bulk propagator is divergence free with respect to the boundary source, and one might have expected that all correlators are also divergence free which would contradict (\ref{mix}). What must happen is that the divergence on the boundary-to-bulk propagator gives a contact term on the boundary, and the resulting divergence of the three point function reduces to the product of two point functions. This is illustrated in the following diagram.

\bigskip

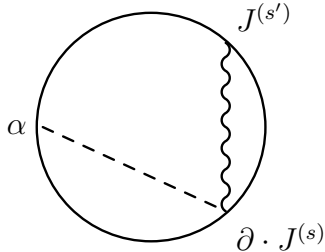
\centerline{\begin{fmffile}{OJJdiv}
        \begin{tabular}{c}
            \begin{fmfgraph*}(40,40)
            \fmfi{plain}{fullcircle scaled .95w shifted (.5w,.5h)}
                \fmfleft{L}
                \fmfright{R}
                \fmffixed{(-.05w,0)}{vO,L}
                \fmffixed{(.19w,0.35h)}{vJ1,R}
                \fmffixed{(.19w,-0.35h)}{vJ2,R}
                \fmf{wiggly}{vJ1,vJ2}
                \fmf{dashes}{vO,vJ1}
                \fmflabel{$\alpha$}{vO}
                \fmflabel{$\partial\cdot J^{(s)}$}{vJ1}
                \fmflabel{$J^{(s')}$}{vJ2}
                \end{fmfgraph*}
        \end{tabular}
        \end{fmffile}
}
\bigskip
\noindent

\section{Four-point functions in the critical $O(N)$ model}

The four point function
\ie
\langle J^{(s_1)}(x_1, \varepsilon_1)J^{(s_2)}(x_2, \varepsilon_2)J^{(s_3)}(x_3, \varepsilon_3)J^{(s_4)}(x_4, \varepsilon_4) \rangle
\fe
can be calculated in $1/N$ expansion, as explained in \cite{LangRuhl}. We will focus on the {\it difference} between this four point function and the corresponding four-point function of conserved currents in the free $O(N)$ vector theory. At leading order in $1/N$, we have
\ie
&\Delta \langle J^{(s_1)} J^{(s_2)} J^{(s_3)} J^{(s_4)}  \rangle \equiv \langle J^{(s_1)} J^{(s_2)} J^{(s_3)} J^{(s_4)}  \rangle_{critical} - \langle J^{(s_1)} J^{(s_2)} J^{(s_3)} J^{(s_4)}  \rangle_{free}
\\
&= \int d^3 y d^3 z \langle J^{(s_1)} J^{(s_2)} \A(y)\rangle D^{-1}_\A(y,z) \langle \A(z) J^{(s_3)} J^{(s_4)}  \rangle + (2\leftrightarrow 3) + (2\leftrightarrow 4)
\fe
where $D^{-1}_\A(y,z)$ is the inverse propagator for the Lagrangian multiplier field $\A$ in position space, obtained from integrating out $\phi_i$ at one-loop. The RHS are expressed in terms of three point functions in the critical $O(N)$ model. In momentum space, we have (still suppressing the polarization vectors)
\ie\label{fac}
&\left. \Delta\langle J^{(s_1)}(p_1) J^{(s_2)}(p_2) J^{(s_3)}(p_3) J^{(s_4)}(p_4)  \rangle\right|_{p_1+p_2+p_3+p_4=0}
\\
&= -{1\over |p_1+p_2|}\langle J^{(s_1)}(p_1) J^{(s_2)}(p_2) \A(-p_1-p_2)\rangle \langle \A(p_1+p_2) J^{(s_3)}(p_3) J^{(s_4)}(p_4) \rangle + (2\leftrightarrow 3) + (2\leftrightarrow 4)\\
\fe
In the next section, we will see that this structure arises naturally in the bulk higher spin gauge theory.

\bigskip

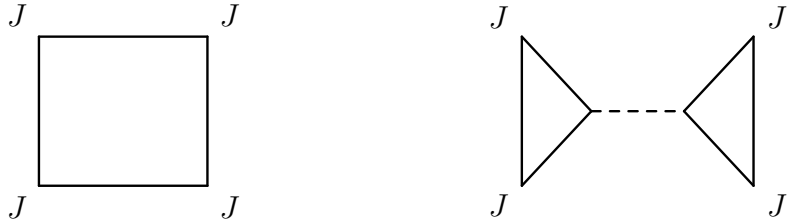
\centerline{\begin{fmffile}{JJJJfree}
        \begin{tabular}{c}
            \begin{fmfgraph*}(40,40)
                \fmfleft{L}
                \fmfright{R}
                \fmffixed{(-.15w,0.31h)}{vJ3,L}
                \fmffixed{(-.15w,-0.31h)}{vJ4,L}
                \fmffixed{(.15w,0.31h)}{vJ1,R}
                \fmffixed{(.15w,-0.31h)}{vJ2,R}
                \fmffixed{(-.32w,0)}{v1,L}
                \fmffixed{(-.68w,0)}{v2,L} 
                \fmffixed{(-.5w,-0.01h)}{t,L}
                \fmf{plain}{vJ1,vJ2,vJ4,vJ3,vJ1}
                \fmflabel{$J$}{vJ1}
                \fmflabel{$J$}{vJ2}
                \fmflabel{$J$}{vJ3}
                \fmflabel{$J$}{vJ4}
             \end{fmfgraph*}
        \end{tabular}
        \end{fmffile}
~~~~~~~~~~~~~~
\begin{fmffile}{JJJJcrit}
        \begin{tabular}{c}
            \begin{fmfgraph*}(55,40)
                \fmfleft{L}
                \fmfright{R}
                \fmffixed{(-.15w,0.31h)}{vJ3,L}
                \fmffixed{(-.15w,-0.31h)}{vJ4,L}
                \fmffixed{(.15w,0.31h)}{vJ1,R}
                \fmffixed{(.15w,-0.31h)}{vJ2,R}
                \fmffixed{(-.36w,0)}{v1,L}
                \fmffixed{(-.64w,0)}{v2,L} 
                \fmffixed{(-.5w,-0.01h)}{t,L}
                \fmf{plain}{vJ1,vJ2,v2,vJ1}
                \fmf{plain}{vJ4,vJ3,v1,vJ4}
                \fmf{dashes}{v1,v2}
                \fmflabel{$J$}{vJ1}
                \fmflabel{$J$}{vJ2}
                \fmflabel{$J$}{vJ3}
                \fmflabel{$J$}{vJ4}
             \end{fmfgraph*}
        \end{tabular}
        \end{fmffile} }  
        \bigskip     
\centerline{Diagrams that contribute at leading order in $1/N$ to the four-point function.}
\bigskip
\noindent

\section{Four-point functions from higher spin gauge theory in $AdS_4$}

Vasiliev's minimal bosonic higher spin gauge theory in $AdS_4$ with the ``standard" $\Delta=1$ boundary condition on the bulk scalar $\varphi$ is believed to be dual to the free $O(N)$ vector theory, whereas the same bulk theory with $\Delta=2$ boundary condition on $\varphi$ is expected to be dual to the critical $O(N)$ model. In perturbation theory, the boundary condition affects correlation functions only through a modification of the bulk scalar propagator \cite{Aharony:1999ti, D'Hoker:2002aw},\footnote{For general mass $m$ the bulk scalar propagator is written in terms of the confluent hypergeometric function. In the special case of $m^2=-2/R^2$, the expression reduces to elementary functions.}
\ie
G_\Delta(x,x') = {1\over 4\pi^2}{\xi^\Delta\over 1-\xi^2},~~~~\xi = {1\over \cosh d(x,x')},
\fe
where $d(x,x')$ is the geodesic distance between $x$ and $x'$, and $\Delta=1$ or $2$ is the dimension of the dual scalar operator. In Poincar\'e coordinates $(\vec x,z)$, where the $AdS_4$ metric is written as
\ie
ds^2 = {dz^2 + d\vec x^2\over z^2},
\fe
we have
\ie
\xi = {2zz'\over (\vec x-\vec x')^2+z^2+z'^2}.
\fe
The nonlinear bulk equation of motion for the scalar takes the form
\ie
(\Box-m^2) \varphi(x) = {\cal J}(x).
\fe
where ${\cal J}(x) = {\cal J}^{(2)}(x) + {\cal J}^{(3)}(x)+\cdots$ is quadratic and higher order in bulk fields of all spins. The {\it difference} between the boundary four-point function of the $\Delta=1$ and $\Delta=2$ boundary condition,
\ie
\Delta \langle J^{(s_1)} J^{(s_2)} J^{(s_3)} J^{(s_4)}  \rangle ,
\fe
receives the contribution from a scalar intermediate channel only, and can be computed as
\ie
& \Delta \langle J^{(s_1)}(\vec x_1,\varepsilon_1) J^{(s_2)}(\vec x_2,\varepsilon_2) J^{(s_3)}(\vec x_3,\varepsilon_3) J^{(s_4)}(\vec x_4,\varepsilon_4)  \rangle
\\
&= \int d^4x \sqrt{g(x)} \int d^4x'\sqrt{g(x')} \big[G_{\Delta=2}(x,x') - G_{\Delta=1}(x,x') \big]
\\
&~~~\times
{\cal J}^{(s_1,s_2)}(x|\vec x_1,\varepsilon_1, \vec x_2,\varepsilon_2) 
{\cal J}^{(s_3,s_4)}(x'|\vec x_3,\varepsilon_3, \vec x_4,\varepsilon_4)  + (2\leftrightarrow 3) + (2\leftrightarrow 4).
\fe
Here ${\cal J}^{(s_1,s_2)}(x|\vec x_1,\varepsilon_1, \vec x_2,\varepsilon_2)$ for instance is defined as the variation of the quadratic part ${\cal J}^{(2)}(x)$ of ${\cal J}(x)$, evaluated on the solution of the linearized bulk higher spin equations of motion, and varied with respect to the boundary sources for the spin $s_i$ field at $x_i$ with polarization vector $\varepsilon_i$, $i=1,2$. In particular,
\ie
\int d^4 x\sqrt{g} K_\Delta(x;\vec x_0) {\cal J}^{(s_1,s_2)}(x|\vec x_1,\varepsilon_1, \vec x_2,\varepsilon_2) 
\fe
where $K_\Delta(x;\vec x_0)$ is the boundary to bulk propagator for the scalar $\varphi$, gives the tree level three point function
\ie
\langle {\cal O}_\Delta(x_0) J^{(s_1)}(x_1,\varepsilon_1) J^{(s_2)}(x_2,\varepsilon_2) \rangle_{free}.
\fe
We will now Fourier transform the correlators to their momentum space expressions, and write $G_\Delta(p;z,z')$ the bulk scalar propagator after Fourier transforming $\vec x, \vec x'$ (but not $z,z'$), and similarly $K_\Delta(p;z)$ the Fourier transformed boundary to bulk propagator. The structure (\ref{fac}) will hold if the following factorization property holds for the difference of the bulk propagator of two different boundary conditions,
\ie\label{gid}
G_{\Delta=2}(p;z,z') - G_{\Delta=1}(p;z,z') = -{1\over |p|} K_{\Delta=2}(p;z) K_{\Delta=2}(p;z')
\fe
Using
\ie
&G_{\Delta=2}(x,x') - G_{\Delta=1}(x,x') = -{1\over 2\pi^2}{zz'\over (\vec x-\vec x')^2 + (z+z')^2},
\\
&G_{\Delta=2}(p;z,z') - G_{\Delta=1}(p;z,z') =- {zz'\over |p|} e^{-|p|(z+z')},
\fe
and
\ie
& K_{\Delta=2}(x;\vec x_0) = {1\over\pi^2} \left[ {z\over (\vec x-\vec x_0)^2+z^2} \right]^2,
\\
& K_{\Delta=2}(p;z) = ze^{-|p|z},
\fe
(\ref{gid}) is easily verified. This shows that the four point function computed from the bulk theory with $\Delta=2$ boundary condition indeed knows the intermediate $\A$ channel contribution of the critical $O(N)$ vector model. Note that our derivation here does not rely on the details of interactions in Vasiliev's theory, but only the structure of bulk scalar propagators. The structure we find here is somewhat reminiscent of \cite{Raju-BCFW}.

\bigskip

\begin{fmffile}{JJOJJ}
        \begin{tabular}{c}
            \begin{fmfgraph*}(40,40)
            \fmfi{plain}{fullcircle scaled .95w shifted (.5w,.5h)}
                \fmfleft{L}
                \fmfright{R}
                \fmffixed{(-.15w,0.31h)}{vJ3,L}
                \fmffixed{(-.15w,-0.31h)}{vJ4,L}
                \fmffixed{(.15w,0.31h)}{vJ1,R}
                \fmffixed{(.15w,-0.31h)}{vJ2,R}
                \fmffixed{(-.32w,0)}{v1,L}
                \fmffixed{(-.68w,0)}{v2,L} 
                \fmffixed{(-.5w,-0.01h)}{t,L}
                \fmf{wiggly}{vJ1,v2,vJ2}
                \fmf{wiggly}{vJ3,v1,vJ4}
                \fmf{dashes}{v1,v2}
                \fmflabel{$\Delta=2$}{t}
                \fmffixed{(0,0.02h)}{t2,t}
                \fmflabel{$p$}{t2}
             \end{fmfgraph*}
        \end{tabular}
$-\!\!\!\!\!$
        \end{fmffile}
        \begin{fmffile}{JJOJJ2}
        \begin{tabular}{c}
            \begin{fmfgraph*}(40,40)
            \fmfi{plain}{fullcircle scaled .95w shifted (.5w,.5h)}
                \fmfleft{L}
                \fmfright{R}
                \fmffixed{(-.15w,0.31h)}{vJ3,L}
                \fmffixed{(-.15w,-0.31h)}{vJ4,L}
                \fmffixed{(.15w,0.31h)}{vJ1,R}
                \fmffixed{(.15w,-0.31h)}{vJ2,R}
                \fmffixed{(-.32w,0)}{v1,L}
                \fmffixed{(-.68w,0)}{v2,L} 
                \fmffixed{(-.5w,-0.01h)}{t,L}
                \fmf{wiggly}{vJ1,v2,vJ2}
                \fmf{wiggly}{vJ3,v1,vJ4}
                \fmf{dashes}{v1,v2}
                \fmflabel{$\Delta=1$}{t}
                \fmffixed{(0,0.02h)}{t2,t}
                \fmflabel{$p$}{t2}
             \end{fmfgraph*}
        \end{tabular}
        \end{fmffile}\\
$~~~~~~~~=~~ -|p|^{-1}~\times\!\!\!\!\!$
\begin{fmffile}{JJOJJcut}
        \begin{tabular}{c}
            \begin{fmfgraph*}(40,40)
            \fmfi{plain}{fullcircle scaled .95w shifted (.5w,.5h)}
                \fmfleft{L}
                \fmfright{R}
                \fmffixed{(-.15w,0.31h)}{vJ3,L}
                \fmffixed{(-.15w,-0.31h)}{vJ4,L}
                \fmffixed{(.15w,0.31h)}{vJ1,R}
                \fmffixed{(.15w,-0.31h)}{vJ2,R}
                \fmffixed{(.55w,0.49h)}{w1,R}
                \fmffixed{(.45w,0.49h)}{w2,R}
                \fmffixed{(-.25w,0)}{v1,L}
                \fmffixed{(-.75w,0)}{v2,L}
                \fmf{wiggly}{vJ1,v2,vJ2}
                \fmf{wiggly}{vJ3,v1,vJ4}
                \fmf{dashes}{v1,w1}
                \fmf{dashes}{v2,w2}
                \fmffixed{(-.05w,0)}{t,w1}
                \fmflabel{$\Delta=2$}{t}
                \fmffixed{(0,-0.4h)}{t1,w1}
                \fmffixed{(0,-0.4h)}{t2,w2}
                \fmflabel{$p$}{t1}
                \fmflabel{$p$}{t2}
             \end{fmfgraph*}
        \end{tabular}
        \end{fmffile}
        $=~~ -|p|~\times\!\!\!\!\!$
\begin{fmffile}{JJOJJcut2}
        \begin{tabular}{c}
            \begin{fmfgraph*}(40,40)
            \fmfi{plain}{fullcircle scaled .95w shifted (.5w,.5h)}
                \fmfleft{L}
                \fmfright{R}
                \fmffixed{(-.15w,0.31h)}{vJ3,L}
                \fmffixed{(-.15w,-0.31h)}{vJ4,L}
                \fmffixed{(.15w,0.31h)}{vJ1,R}
                \fmffixed{(.15w,-0.31h)}{vJ2,R}
                \fmffixed{(.55w,0.49h)}{w1,R}
                \fmffixed{(.45w,0.49h)}{w2,R}
                \fmffixed{(-.25w,0)}{v1,L}
                \fmffixed{(-.75w,0)}{v2,L}
                \fmf{wiggly}{vJ1,v2,vJ2}
                \fmf{wiggly}{vJ3,v1,vJ4}
                \fmf{dashes}{v1,w1}
                \fmf{dashes}{v2,w2}
                \fmffixed{(-.05w,0)}{t,w1}
                \fmflabel{$\Delta=1$}{t}
                \fmffixed{(0,-0.4h)}{t1,w1}
                \fmffixed{(0,-0.4h)}{t2,w2}
                \fmflabel{$p$}{t1}
                \fmflabel{$p$}{t2}
             \end{fmfgraph*}
        \end{tabular}
        \end{fmffile}
        \\
        \\
        \\
\centerline{``Cutting" the bulk four-point function by means of the identity (\ref{gid}).}

\section{A general argument for $n$ point functions}

To begin with, consider an $n$-point function of higher spin currents in the critical $O(N)$ model, without any scalar operator, written in momentum space as
\ie
\langle J_1(p_1)\cdots J_n(p_n)\rangle.
\fe
Denote by ${\cal G}$ be a bulk $\ell$-loop Witten graph, and by $\langle {\cal G}\rangle_{\Delta=2}$ its contribution to the $n$-point boundary correlator with $\Delta=2$ boundary condition. Let ${\cal I}$ be the index set labelling all internal scalar lines in ${\cal G}$. For each subset ${\cal I'}\subset {\cal I}$, Let ${\cal G}_{\cal I'}(\{k_i^{(1)},k_i^{(2)}\}_{i\in {\cal I}'})$ be the Witten graph obtained by cutting open all scalar lines in ${\cal I'}$, and replace each cut scalar line, say the one labelled by $i\in {\cal I}'$, with a pair of external scalar lines with $\Delta=1$ boundary condition and momenta $k_i^{(1)}$, $k_i^{(2)}$. 

\bigskip

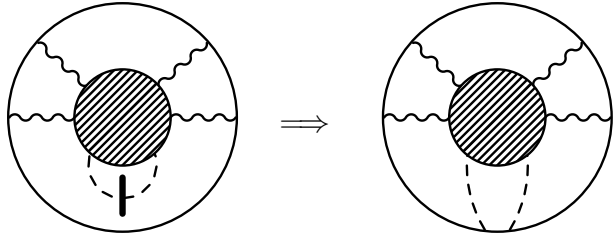
\centerline{\begin{fmffile}{beforecut}
        \begin{tabular}{c}
            \begin{fmfgraph*}(40,40)
            \fmfi{plain}{fullcircle scaled .95w shifted (.5w,.5h)}
                \fmfleft{L}
                \fmfright{R}
                \fmffixed{(-.15w,-0.31h)}{vJ4,L}
                \fmffixed{(.15w,-0.31h)}{vJ2,R}
                \fmffixed{(.03w,0)}{vJ5,R}
                \fmffixed{(-.03w,0)}{vJ6,L}
                \fmffixed{(-.5w,0)}{v,L} 
                \fmfblob{.4w}{v}
                \fmffixed{(-.12w,0.12h)}{w1,v}
                \fmffixed{(.12w,0.12h)}{w2,v}
                \fmffixed{(0,0.25h)}{c1,v}
                \fmffixed{(0,0.4h)}{c2,v}
                \fmffixed{(-.5w,-0.01h)}{t,L}
                \fmf{wiggly}{vJ2,v,vJ4}
                \fmf{wiggly}{vJ5,v,vJ6}
                \fmf{dashes,left=1.8}{w1,w2}
                \fmf{plain,width=2.5}{c1,c2}
                \end{fmfgraph*}
        \end{tabular}
        \end{fmffile}
        $\Longrightarrow$
\begin{fmffile}{aftercut}
        \begin{tabular}{c}
            \begin{fmfgraph*}(40,40)
            \fmfi{plain}{fullcircle scaled .95w shifted (.5w,.5h)}
                \fmfleft{L}
                \fmfright{R}
                \fmffixed{(-.15w,-0.31h)}{vJ4,L}
                \fmffixed{(.15w,-0.31h)}{vJ2,R}
                \fmffixed{(.03w,0)}{vJ5,R}
                \fmffixed{(-.03w,0)}{vJ6,L}
                \fmffixed{(-.5w,0)}{v,L} 
                \fmfblob{.4w}{v}
                \fmffixed{(-.12w,0.12h)}{w1,v}
                \fmffixed{(.12w,0.12h)}{w2,v}
                \fmffixed{(-0.03w,0.5h)}{c1,v}
                \fmffixed{(0.03w,0.5h)}{c2,v}
                \fmffixed{(-.5w,-0.01h)}{t,L}
                \fmf{wiggly}{vJ2,v,vJ4}
                \fmf{wiggly}{vJ5,v,vJ6}
                \fmf{dashes,left=.2}{w1,c1}
                \fmf{dashes,right=.2}{w2,c2}
                \end{fmfgraph*}
        \end{tabular}
        \end{fmffile}
        }
        \bigskip
\centerline{Cutting procedure: the difference between $\Delta=2$ and $\Delta=1$ bulk propagators}\centerline{is replaced by the product of two propagators to the boundary.}

\bigskip

Now using
\ie
G_{\Delta=2}(q;z,z') - G_{\Delta=1}(q;z,z') = -|q| K_{\Delta=1}(q,z) K_{\Delta=1}(-q,z'),
\fe
we can write
\ie\label{dec}
\langle {\cal G}\rangle_{\Delta=2} = \sum_{{\cal I'}\subset {\cal I}} \int\prod_{i\in {\cal I'}} d^3 q_i \,(-|q_i|)\left\langle {\cal G}_{\cal I'}(\{k_i^{(1)}=q_i,k_i^{(2)}=-q_i\}_{i\in {\cal I}'})\right\rangle_{\Delta=1}.
\fe
where on the RHS, $\langle {\cal G}_{\cal I'} \rangle_{\Delta=1}$ is evaluated as a Witten diagram with $\Delta=1$ boundary condition (all internal scalar lines are replaced by $G_{\Delta=1}$ as well). In writing the above, a delta function imposing momentum conservation is included in each connected correlation function, and the integration over $q_i$ may involve nontrivial loop integrals after the delta functions are integrated out. The key observation here is that the $1/N$ diagrammatic expansion of the critical $O(N)$ model admits a decomposition into diagrams for Wick contractions of currents in the free theory, sewed together by $\A$-propagators in essentially the same way. If we assume that the duality holds with $\Delta=1$ boundary condition, namely the sum of all Witten diagrams $\langle {\cal G} \rangle_{\Delta=1}$ with external legs $J_1(p_1),\cdots, J_n(p_n)$ produces the correct $n$-point function of the free $O(N)$ theory, then the $n$-point functions of higher spin currents of the critical $O(N)$ model is precisely reproduced by summing over $\langle {\cal G} \rangle_{\Delta=2}$, by virtue of (\ref{dec}).

The four-point function discussed in the previous sections is a special case of this construction. This cutting procedure works for loop diagrams as well, and relates the difference between loops of $\Delta=2$ and $\Delta=1$ scalar propagators to diagrams in which the loop is cut open and replaced by two external scalar lines. Note that $G_{\Delta=2}-G_{\Delta=1}$ is free of short distance singularity, and we have assumed that the UV divergences cancel among loop diagrams in Vasiliev theory with $\Delta=1$ boundary condition, due to higher spin symmetry, which is necessary for the vanishing of $1/N$ corrections to correlators in the free $O(N)$ theory. It is also straightforward to generalize the above construction to include the case where a number of scalar operators $\A$ are inserted into the correlation function. 

Let us illustrate this further with the example of bulk one-loop correction to the two-point function $\langle JJ\rangle$ of a higher spin current $J$. The bulk one-loop diagrams involving at least a scalar propagator give different contributions in the case of $\Delta=2$ boundary condition as opposed to $\Delta=1$ boundary condition. These diagrams are listed below.

\bigskip
        
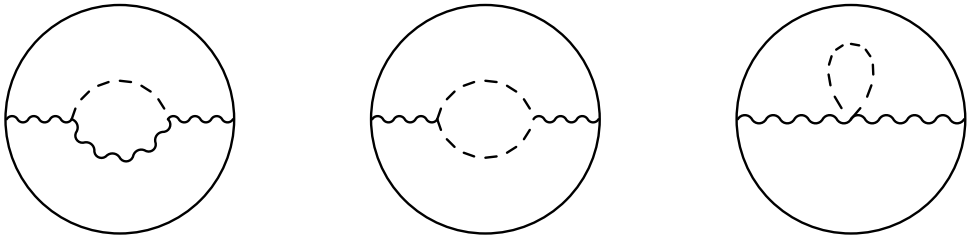
\begin{fmffile}{JJloop1}
        \begin{tabular}{c}
            \begin{fmfgraph*}(40,40)
            \fmfi{plain}{fullcircle scaled .95w shifted (.5w,.5h)}
                \fmfleft{L}
                \fmfright{R}
                \fmffixed{(-.03w,0)}{vL,L}
                \fmffixed{(.03w,0)}{vR,R}
                \fmffixed{(-.3w,0)}{v1,L}
                \fmffixed{(-.7w,0)}{v2,L}
                \fmf{dashes,left=.8}{v1,v2}
                \fmf{wiggly,right=.8}{v1,v2}
                \fmf{wiggly}{vL,v1}
                \fmf{wiggly}{v2,vR}
             \end{fmfgraph*}
        \end{tabular}
        \end{fmffile}
        ~~~~
\begin{fmffile}{JJloop2}
        \begin{tabular}{c}
            \begin{fmfgraph*}(40,40)
            \fmfi{plain}{fullcircle scaled .95w shifted (.5w,.5h)}
                \fmfleft{L}
                \fmfright{R}
                \fmffixed{(-.03w,0)}{vL,L}
                \fmffixed{(.03w,0)}{vR,R}
                \fmffixed{(-.3w,0)}{v1,L}
                \fmffixed{(-.7w,0)}{v2,L}
                \fmf{dashes,left=.8}{v1,v2}
                \fmf{dashes,right=.8}{v1,v2}
                \fmf{wiggly}{vL,v1}
                \fmf{wiggly}{v2,vR}
             \end{fmfgraph*}
        \end{tabular}
        \end{fmffile}
        ~~~~
\begin{fmffile}{JJloop3}
        \begin{tabular}{c}
            \begin{fmfgraph*}(40,40)
            \fmfi{plain}{fullcircle scaled .95w shifted (.5w,.5h)}
                \fmfleft{L}
                \fmfright{R}
                \fmffixed{(-.03w,0)}{vL,L}
                \fmffixed{(.03w,0)}{vR,R}
                \fmffixed{(-.5w,0)}{v,L}
                \fmf{dashes}{v,v}
                \fmf{wiggly}{vL,v,vR}
             \end{fmfgraph*}
        \end{tabular}
        \end{fmffile}

\noindent Here we have assumed some appropriate gauge fixing and ghost contributions, which do not affect our argument in relating the $\Delta=1$ and $\Delta=2$ correlators. We have omitted tadpole diagrams so far, which are a priori included in the cutting argument above. Nonetheless, the tadpole diagrams should vanish by themselves, for the following reason. While the tadpoles for higher spin fields clearly vanish by symmetry, the tadpole for the bulk scalar in the $\Delta=1$ theory must also vanish provided that the equation of motion is not renormalized, due to higher spin gauge symmetry. Changing from $\Delta=1$ to $\Delta=2$ boundary condition does not shift the tadpole for the bulk scalar; this is related to the vanishing of $\A$ tadpole in the critical $O(N)$ model, which amounts to tuning to criticality.\footnote{Note however that the bulk diagrams, after cutting, are not in one-to-one correspondence with Feynman diagrams for the $1/N$ expansion of the critical $O(N)$ model by cutting $\A$-propagators. Rather, it is the sum of all bulk diagrams at a given order, with the same external lines and $\Delta=1$ internal scalar propagators, that agrees with the sum of appropriate diagrams of free Wick contractions in the boundary theory.}

The following is an example of cutting one internal scalar line, from which we obtain a four-point tree diagram with two $J$'s and two scalar operators on the boundary. The remaining, uncut, internal propagator involves either the scalar or higher spin fields.

\bigskip
        
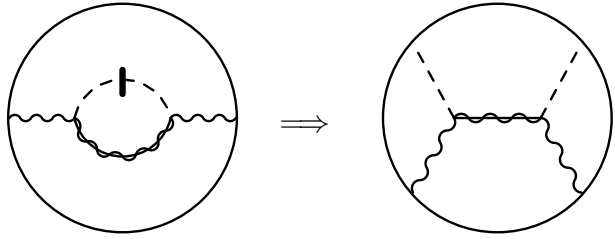
\centerline{\begin{fmffile}{JJloop1cut}
        \begin{tabular}{c}
            \begin{fmfgraph*}(40,40)
            \fmfi{plain}{fullcircle scaled .95w shifted (.5w,.5h)}
                \fmfleft{L}
                \fmfright{R}
                \fmffixed{(-.03w,0)}{vL,L}
                \fmffixed{(.03w,0)}{vR,R}
                \fmffixed{(-.3w,0)}{v1,L}
                \fmffixed{(-.7w,0)}{v2,L}
                \fmffixed{(-.5w,-.1h)}{c1,L}
                \fmffixed{(-.5w,-.2h)}{c2,L}
                \fmf{dashes,left=.8}{v1,v2}
                \fmf{wiggly,right=.8}{v1,v2}
                \fmf{plain,right=.8}{v1,v2}
                \fmf{plain,width=2.5}{c1,c2}
                \fmf{wiggly}{vL,v1}
                \fmf{wiggly}{v2,vR}
             \end{fmfgraph*}
        \end{tabular}
        \end{fmffile}
$\Longrightarrow$
\begin{fmffile}{JJloop1cut2}
        \begin{tabular}{c}
            \begin{fmfgraph*}(40,40)
            \fmfi{plain}{fullcircle scaled .95w shifted (.5w,.5h)}
                \fmfleft{L}
                \fmfright{R}
                \fmffixed{(-.15w,0.31h)}{vJ3,L}
                \fmffixed{(-.15w,-0.31h)}{vJ4,L}
                \fmffixed{(.15w,0.31h)}{vJ1,R}
                \fmffixed{(.15w,-0.31h)}{vJ2,R}
                \fmffixed{(-.32w,0)}{v1,L}
                \fmffixed{(-.68w,0)}{v2,L} 
                \fmffixed{(-.5w,-0.01h)}{t,L}
                \fmf{wiggly}{vJ1,v2}
                \fmf{dashes}{v2,vJ2}
                \fmf{wiggly}{vJ3,v1}
                \fmf{dashes}{v1,vJ4}
                \fmf{plain}{v1,v2}
                \fmf{wiggly}{v1,v2}
             \end{fmfgraph*}
        \end{tabular}
        \end{fmffile}
        }
\noindent When the other internal propagator is also a scalar line, it was a $\Delta=2$ propagator to begin with. In reducing it to a $\Delta=1$ propagator, one obtains an additional contribution that is represented by cutting this scalar line as well. The result is a product of two three point functions in this case.    

\bigskip    
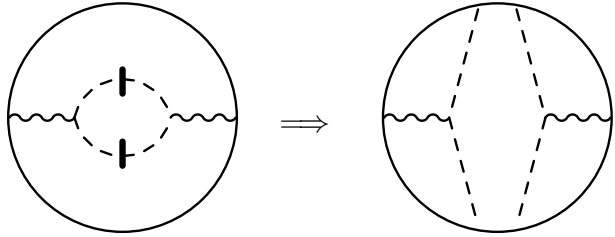
\centerline{\begin{fmffile}{JJloop2cut}
        \begin{tabular}{c}
            \begin{fmfgraph*}(40,40)
            \fmfi{plain}{fullcircle scaled .95w shifted (.5w,.5h)}
                \fmfleft{L}
                \fmfright{R}
                \fmffixed{(-.03w,0)}{vL,L}
                \fmffixed{(.03w,0)}{vR,R}
                \fmffixed{(-.3w,0)}{v1,L}
                \fmffixed{(-.7w,0)}{v2,L}
                \fmffixed{(-.5w,-.1h)}{c1,L}
                \fmffixed{(-.5w,-.2h)}{c2,L}
                \fmffixed{(-.5w,.1h)}{c3,L}
                \fmffixed{(-.5w,.2h)}{c4,L}
                \fmf{dashes,left=.8}{v1,v2}
                \fmf{dashes,right=.8}{v1,v2}
                \fmf{plain,width=2.5}{c1,c2}
                \fmf{plain,width=2.5}{c3,c4}
                \fmf{wiggly}{vL,v1}
                \fmf{wiggly}{v2,vR}
             \end{fmfgraph*}
        \end{tabular}
        \end{fmffile}
$\Longrightarrow$
\begin{fmffile}{JJloop2cut2}
        \begin{tabular}{c}
            \begin{fmfgraph*}(40,40)
            \fmfi{plain}{fullcircle scaled .95w shifted (.5w,.5h)}
                \fmfleft{L}
                \fmfright{R}
                \fmffixed{(-.03w,0)}{vL,L}
                \fmffixed{(.03w,0)}{vR,R}
                \fmffixed{(-.3w,0)}{v1,L}
                \fmffixed{(-.7w,0)}{v2,L}
                \fmffixed{(-.42w,-.45h)}{c1,L}
                \fmffixed{(-.58w,-.45h)}{c2,L}
                \fmffixed{(-.42w,.45h)}{c3,L}
                \fmffixed{(-.58w,.45h)}{c4,L}
                \fmf{dashes}{c1,v1,c3}
                \fmf{dashes}{c2,v2,c4}
                \fmf{wiggly}{vL,v1}
                \fmf{wiggly}{v2,vR}
             \end{fmfgraph*}
        \end{tabular}
        \end{fmffile}
        }
\bigskip

In the critical $O(N)$ model, the diagrams that give rise to the first $1/N$ correction to the two point function $\langle JJ\rangle$ are listed below. The contributions from graphs $(a), (b)$ {\sl altogether} is reproduced by cutting one internal scalar line of the bulk one-loop diagrams, as explained above. $(c)$ is reproduced by the bulk contribution from cutting two internal scalar lines.


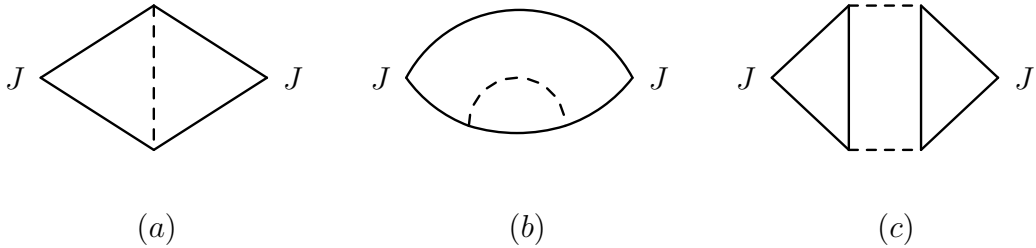
\begin{fmffile}{JJcrit1}
        \begin{tabular}{c}
            \begin{fmfgraph*}(40,40)
                \fmfleft{L}
                \fmfright{R}
                \fmffixed{(-.03w,0)}{vL,L}
                \fmffixed{(.03w,0)}{vR,R}
                \fmffixed{(0,-.6h)}{v1,v2}
                \fmf{plain}{vL,v1,vR}
                \fmf{plain}{vL,v2,vR}
                \fmf{dashes,tension=0}{v1,v2}
                \fmflabel{$J$}{vL}
                \fmflabel{$J$}{vR}
             \end{fmfgraph*}
        \end{tabular}
        \end{fmffile}
        ~~~~
\begin{fmffile}{JJcrit2}
        \begin{tabular}{c}
            \begin{fmfgraph*}(40,40)
                \fmfleft{L}
                \fmfright{R}
                \fmffixed{(-.03w,0)}{vL,L}
                \fmffixed{(.03w,0)}{vR,R}
                \fmffixed{(.4w,0)}{v1,v2}
                \fmffixed{(-.26w,.2h)}{v1,vL}
                \fmf{plain,left=.6}{vL,vR}
                \fmf{plain,right=.15}{vL,v1,v2,vR}
                \fmf{dashes,left=1,tension=0}{v1,v2}
                \fmflabel{$J$}{vL}
                \fmflabel{$J$}{vR}
             \end{fmfgraph*}
        \end{tabular}
        \end{fmffile}
        ~~~~
\begin{fmffile}{JJcrit3}
        \begin{tabular}{c}
            \begin{fmfgraph*}(40,40)
                \fmfleft{L}
                \fmfright{R}
                \fmffixed{(-.03w,0)}{vL,L}
                \fmffixed{(.03w,0)}{vR,R}
                \fmffixed{(0,-.6h)}{v1,v2}
                \fmffixed{(.3w,0)}{v1,v3}
                \fmffixed{(0,-.6h)}{v3,v4}
                \fmf{plain}{vL,v1,v2,vL}
                \fmf{plain}{vR,v3,v4,vR}
                \fmf{dashes,tension=0}{v1,v3}
                \fmf{dashes,tension=0}{v2,v4}
                \fmflabel{$J$}{vL}
                \fmflabel{$J$}{vR}
             \end{fmfgraph*}
        \end{tabular}
        \end{fmffile}
        \\
{$~~~~~~~~~~~~~~~~~(a)~~~~~~~~~~~~~~~~~~~~~~~~~~~~~~~~(b)~~~~~~~~~~~~~~~~~~~~~~~~~~~~~~~~(c)$}
\bigskip

Our argument also implies, in particular, that the $1/N$ contributions to the anomalous dimensions of the higher spin currents in the critical $O(N)$ model, which can be computed through the loop corrections to the two-point functions, are indeed correctly produced by the bulk loop computation, assuming that the duality with free $O(N)$ theory holds in the case of $\Delta=1$ boundary condition.

\section{Concluding remarks}

The equations of motion of the (parity invariant) Vasiliev system in $AdS_4$ \cite{Vasiliev:1999ba} are highly constrained by higher spin gauge symmetries and is conceivably not renormalized with $\Delta=1$ boundary condition.\footnote{In this paper we have restricted our discussion to the type A minimal bosonic theory, in which the bulk scalar is parity even \cite{Vasiliev:1999ba, Sezgin:2003pt}. In the type B theory where the bulk scalar is parity odd, the boundary condition that preserves the higher spin symmetry assigns $\Delta=2$ to the scalar operator. In the more general parity violating higher spin gauge theories of \cite{Vasiliev:1999ba}, generally, neither boundary condition preserves higher spin symmetry. This will be explained in \cite{GMPTY}.} Assuming that the bulk tree level diagrams reproduces the correct $n$ point functions of the free $O(N)$ theory, and that all loop corrections cancel with $\Delta=1$ boundary condition, our argument then shows that the theory with $\Delta=2$ boundary condition has a (UV finite) perturbative expansion, which order by order matches the $1/N$ expansion of the critical $O(N)$ vector model (where the loops are built using $\A$ propagators).

While the higher spin symmetry is broken by the $\Delta=2$ boundary condition, this breaking is controlled by the bulk coupling constant (or $1/N$), and the anomalous dimensions of the boundary higher spin currents are suppressed by $1/N$. Ultimately, one would be interested in bulk theories in which the masses of the higher spin fields can be lifted while keeping the gravity coupling weak. Though it is unclear how to do this within Vasiliev's framework, which may require coupling the higher spin gauge fields to matter fields in some way, we may suspect that a UV finite higher spin gauge theory could be a useful starting point to understand quantum gravity theories with a standard semi-classical gravity limit.

\subsection*{Acknowledgments}

We would like to thank Suvrat Raju for inspiring conversations.
X.Y. would like to thank the hospitality of Perimeter Institute during the course of this work.
S.G. is supported by Perimeter Institute
for Theoretical Physics. Research at Perimeter Institute is supported by the
Government of Canada through Industry Canada and by the Province of Ontario through
the Ministry of Research $\&$ Innovation. X.Y. is supported in part by the Fundamental Laws Initiative Fund at Harvard University, and by NSF Award PHY-0847457.

\appendix

\section{An example of higher spin symmetry breaking in the three point function}

We have seen that the three point functions of the scalar operator and two higher spin currents, $\langle J^{(0)} J^{(s)} J^{(s')}\rangle$, at leading order in $1/N$ in the free $O(N)$ and critical $O(N)$ vector models,  are related simply by multiplying the propagator $D_\A(p)$ of $\A$ field in momentum space. From the bulk, this was seen as due to the difference in the scalar boundary-to-bulk propagators. When $s$ and $s'$ are different spins, say $s>s'$, we have argued that the three point function $\langle \A\, J^{(s)} J^{(s')}\rangle$ is not conserved with respect to $J^{(s)}$ at {\it leading} order in $1/N$, in the critical theory. One may be puzzled as to why $\langle \A\, \partial\cdot J^{(s)} J^{(s')}\rangle$ is nonzero whereas $\langle {\cal O}\, \partial\cdot J^{(s)} J^{(s')}\rangle$ vanishes in the free theory, since the two are simply related by a factor $D_\A(p)$ in momentum space. This is because the latter is in fact a contact term, and when transformed into momentum space is analytic at zero momenta. 

In the $O(N)$ vector model there are only even spin currents, and the first nontrivial example of a three point function that exhibits higher spin symmetry breaking at leading order in $1/N$ would involve spins 4, 2, and 0. For simplicity, we will consider below the $U(N)$ version of the vector model, and the example of three point function involving currents of spins $s=3$, $s'=1$, and the scalar operator. 

The tensor structure of $\langle J^{(0)} J^{(3)} J^{(1)} \rangle$ is uniquely fixed by conformal symmetry up to normalization, as explained in \cite{Giombi:2011rz}. It is useful though to directly compute $\langle J^{(0)}(-p_1-p_2) J^{(3)}(p_1, \varepsilon_1) J^{(1)}(p_2, \varepsilon_2) \rangle$ in momentum space. WIthout loss of generality, the polarization vectors $\varepsilon_1, \varepsilon_2$ are assumed to be null here. The result is
\ie
\int d^3q {\varepsilon_2\cdot (2q+p_2)\, f_3(\varepsilon_1\cdot q,\varepsilon_1\cdot (p_1-q))\over q^2 (q-p_1)^2 (q+p_2)^2}
\fe
in the free theory, and the same expression multiplied by $-|p_1+p_2|$ in the critical theory. Here $f_3$ is the spin 3 part of the generating function $f(u,v)$ defined in section 2; $f_3(u,v)={1\over 6}(u-v)(u^2-14 uv+v^2)$.

Now taking the divergence on the spin 3 current $J^{(3)}(p_1)$, one obtains
\ie\label{intf}
{1\over 2} \int d^3q\, \varepsilon_2\cdot (2q+p_2)\left[{h(\varepsilon_1\cdot q,\varepsilon_1\cdot (p_1-q))\over (q-p_1)^2 (q+p_2)^2} - {h(\varepsilon_1\cdot (p_1-q),\varepsilon_1\cdot q)\over q^2 (q+p_2)^2}\right]
\fe
in the case of the free theory, where $h(u,v) \equiv u^2-10uv+5v^2$, and the same result multiplied by $-|p_1+p_2|$ in the critical theory. The integral of (\ref{intf}) is the sum of two terms. The first term is analytic at $p_1=0$ or $p_2=0$, when $|p_1+p_2|$ is nonzero; the second term is analytic at $p_1=0$ or $p_1+p_2=0$, when $|p_1|$ is nonzero. Consequently, both give contact terms when Fourier transformed into position space.

If we multiply (\ref{intf}) by $-|p_1+p_2|$, however, as in the critical theory, then we obtain a non-analytic term
\ie
{|p_1+p_2|\over 2}\int d^3 q \,\varepsilon_2\cdot (2q+p_2){h(\varepsilon_1\cdot (p_1-q),\varepsilon_1\cdot q)\over q^2 (q+p_2)^2}
\fe
which factorizes into the product of two point functions $\langle \A(p_1+p_2)\A(-p_1-p_2)\rangle$ and $\langle J^{(1)}(p_2,\varepsilon_1)J^{(1)}(-p_2,\varepsilon_2)\rangle$, with an additional momentum factor.

\end{document}